# Power-law setting of steady concentration in a binary solution within a droplet at diffusion-controlled or free-molecular regimes of the droplet binary growth in the vapor-gas environment

Fedor M. Kuni, Alexandra A. Lezova, and Alexander K. Shchekin

*Department of Statistical Physics, St Petersburg State University, Ulyanovskaya 1, Petrodvoretz, St Petersburg, 198504, Russian Federation*

**Abstract**

The times required for reaching the power-law-in-time growth of the droplet radius after nucleation of a markedly supercritical binary droplet, are found analytically and estimated at diffusion-controlled or free-molecular regimes of isothermal binary vapor condensation. The process of setting steady concentration in the binary solution within the growing droplet at the same regimes of binary condensation has been analyzed. It has been shown that setting the steady droplet concentration has also a power-law character in time. The parameters of the power law are specified for each regime of binary condensation and are linked to thermodynamic and kinetic characteristics of condensing vapors and to the steady concentration established in a binary droplet.

*Keywords*: Aerosol formation; Binary condensation; Droplet growth; Diffusion-controlled regime; Free-molecular regime

# 1. Introduction

The goals of this paper are an analytical investigation of the regularities of setting the steady concentration in the binary solution within a markedly supercritical droplet growing in vapor-gas environment and an analysis of hierarchy of corresponding time scales in the binary condensation process. These goals have a direct relation to the problems of atmospheric aerosol formation. It is assumed that environment consists of two condensing vapors and passive gas-carrier; condensation growth of the droplet is isothermal, and the vapor condensation onto the droplet runs in the continuum diffusion-controlled or kinetic free-molecular regimes. Speaking about a markedly supercritical droplet, we mean that the growth of the droplet is a regular one, and the fluctuation and capillarity effects can be neglected.

The goals formulated above were not raised earlier in the literature on problems of binary condensation and nucleation (Clement et al., 2006; Djikaev et al., 1999; Djikaev et al., 2004; Grinin et al., 2008; Kulmala et al., 1993; Kuni et al., 1998; Kuni & Lezova, 2008; Vesala & Kulmala, 1993). Nevertheless, the significance of these goals have been brought up by works of Grinin et al. (2008) and Kuni & Lezova (2008), where a self-similar solution of the binary condensation problem has been found. The approach developed required a setting of the steady concentration of a solution in a growing droplet, and the questions aroused what are the conditions of such a setting and how fast it occurs. Along with that, the achievement of these goals will rely upon the analytical results obtained by Kulmala et al. (1993) and related to finding the steady concentration in the binary solution within the markedly supercritical droplet growing in the diffusion-controlled regime. These results are important for establishing links between the parameters of the laws of setting the steady concentration in droplet in time, because the parameters themselves depend on the steady concentration.

As a first step, we will find analytically the times between nucleation of a markedly supercritical droplet and achievement of the power-law-in-time growth of the droplet radius at diffusion-controlled or free-molecular regimes of isothermal binary vapor condensation. As a next step, we will derive the power laws of setting steady concentration in time in the binary solution within the growing droplet at the same regimes of binary condensation. It will be shown how the parameters of the power laws are specified for each regime of binary condensation and how they are tied with the thermodynamic and kinetic characteristics of condensing vapors and with the establishing steady concentration in a binary droplet.

## 2. Starting relations of binary condensation onto a droplet

Let $x_1$ and $x_2$ be the numbers of molecules of the first and the second components in the liquid solution within the droplet. We denote as $v_1$ and $v_2$ the partial molecular volumes of these components in the solution. The relative concentrations $c_1$ and $c_2$ of the first and the second components in solution can be written as

$$c_1 \equiv \frac{x_1}{x_1 + x_2}, \quad c_2 \equiv 1 - c_1 \equiv \frac{x_2}{x_1 + x_2} \quad (0 < c_1 < 1, \ 0 < c_2 < 1). \tag{1}$$

Obtaining the equality $(1 - c_1) x_1 = c_1 x_2$ from Eq. (1) and differentiating the equality with respect to time $t$, we have

$$(x_1 + x_2)\frac{dc_1}{dt} = (1 - c_1)\frac{dx_1}{dt} - c_1 \frac{dx_2}{dt}. \tag{2}$$

This equation will be the key one for the following analysis.

As the total volume of the solution in the droplet is the first order homogeneous function with respect to $x_1$ and $x_2$ and the partial volumes $v_1$ and $v_2$ are zero order homogeneous functions with respect to $x_1$ and $x_2$, then

$$v_1 x_1 + v_2 x_2 = 4\pi R^3 / 3 \tag{3}$$

where $R$ is the droplet radius. As follows from Eqs. (1) and (3),

$$x_1 + x_2 = \frac{4\pi R^3}{3} \frac{1}{c_1 v_1 + (1 - c_1) v_2}. \tag{4}$$

Let $n_1$ and $n_2$ be the number densities of molecules of the first and the second components in the vapor-gas environment at the infinite distance from the droplet. We denote as $n_{1\infty}(c_1)$ and $n_{2\infty}(c_2)$ the equilibrium (saturated over liquid solution) values of these densities near the droplet surface (as it was said in Introduction, one can ignore the dependence of these values on the droplet radius $R$ for a markedly supercritical droplet).

We consider the vapor mixture metastable with respect to each component. That means

$$n_1 > n_{1\infty}(c_1), \quad n_2 > n_{2\infty}(c_2). \tag{5}$$

Only under both of conditions (5), the condensation of both vapor components onto droplet will take place.

According to the equations of ideal solution,

$$n_{1\infty}(c_1) = \tilde{n}_{1\infty} c_1, \quad n_{2\infty}(c_2) = \tilde{n}_{2\infty} c_2 \tag{6}$$

where $\tilde{n}_{1\infty}$ and $\tilde{n}_{2\infty}$ are the number densities of molecules of saturated vapors over pure liquids of the first and the second components. For non-ideal solutions, the right-hand sides of relations in Eq. (6) can be represented as the following series in concentrations

$$n_{1\infty}(c_1) = \tilde{n}_{1\infty} c_1 + \sum_{k=2} \alpha_k^{(1)} c_1^k, \quad n_{2\infty}(c_2) = \tilde{n}_{2\infty} c_2 + \sum_{k=2} \alpha_k^{(2)} c_2^k \tag{7}$$

where first terms correspond to the ideal approximation, $\alpha_k^{(1)}$ and $\alpha_k^{(2)}$ ($k = 2,3...$) are the coefficients of the series for the first and second components, respectively.

## 3. Diffusion-controlled regime of droplet growth

Let us consider first binary condensation in a continuum diffusion-controlled regime. The amount of the passive gas in the vapor-gas environment is assumed to be sufficiently large. The isothermal droplet growth runs in the diffusion-controlled regime if the strong inequality $R/\lambda \gg 1$ is valid, where $\lambda$ is the free path length of molecules of the vapors in the passive gas. At that the following equations for numbers of molecules of components in a droplet take place (Fuks, 1958),

$$\frac{dx_1}{dt} = 4\pi D_1 \left[ n_1 - n_{1\infty}(c_1) \right] R, \qquad \frac{dx_2}{dt} = 4\pi D_2 \left[ n_2 - n_{2\infty}(c_2) \right] R \qquad (8)$$

where $D_1$ and $D_2$ are the diffusion coefficients of molecules of vapors in the passive gas.

Let $c_{1s}$ be steady concentration of the first component in the solution within the growing droplet. Taking into account that $dc_{1s}/dt = 0$ and using Eqs. (2) and (8), we find

$$\frac{D_1 \left[ n_1 - n_{1\infty}(c_{1s}) \right]}{D_2 \left[ n_2 - n_{2\infty}(1 - c_{1s}) \right]} = \frac{c_{1s}}{1 - c_{1s}}. \qquad (9)$$

According to the thermodynamic stability conditions of solutions $\partial n_{1\infty}(c_{1s})/\partial c_{1s} > 0$ and $\partial n_{2\infty}(1 - c_{1s})/\partial c_{1s} < 0$ (Landau, Lifshitz, 1976), and the conditions (5), equation (9) unambiguously determines $c_{1s}$ (because the left- and the right-hand sides of Eq. (9) varies in opposite directions at changing $c_{1s}$) and provides $0 < c_{1s} < 1$. Just the equation equivalent to Eq. (9) was derived by Kulmala et al. (1993) for finding concentration $c_{1s}$ at ideality of solution in a growing droplet.

Let us differentiate both sides of Eq. (3) with respect to time $t$ at $c_1 = c_{1s}$ and fixed $\upsilon_1$ and $\upsilon_2$. Using equations (8), we obtain

$$dR^2/dt = \beta^2 \qquad (10)$$

where

$$\beta^2 \equiv 2D_1\upsilon_1 \left[ n_1 - n_{1\infty}(c_{1s}) \right] + 2D_2\upsilon_2 \left[ n_2 - n_{2\infty}(1 - c_{1s}) \right]. \qquad (11)$$

The positivity of the parameter $\beta^2$, that follows from Eqs. (11) and (5), corresponds to droplet growth.

We count time $t$ off the moment $t = 0$ of nucleation of the markedly supercritical droplet. Let $t_0$ and $R_0$ be the values of $t$ and $R$, starting from which the continuum regime of droplet growth (described by Eqs. (8) and (10)) is well established, so that $R_0/\lambda \sim 10 \div 20$. Then the condition $R/\lambda \gg 1$ of the continuum regime will be valid long before the moment of time $t_0$, and, by integrating Eq. (10) with respect to time, we find with a good accuracy that droplet radius satisfies a power law growth in time,

$$R = \beta t^{1/2} \quad (t \geq t_0). \tag{12}$$

All the further consideration in this Section will be referred to times $t \geq t_0$ and radii $R \geq R_0$.

As follows from Eq. (12) at $t = t_0$,

$$t_0 = R_0^2 / \beta^2. \tag{13}$$

The value $t_0$ is the time after nucleation of the markedly supercritical droplet, after which the power law (12) of droplet radius growth comes into force. With the help of Eq. (13) we can rewrite Eq. (12) as

$$R / R_0 = (t / t_0)^{1/2} \quad (t \geq t_0). \tag{14}$$

Now we are interested in setting of the steady concentration $c_{1s}$. With the help of Eq. (2), we can formulate an iteration method to finding the deviation of concentration $c_1$ from the steady concentration $c_{1s}$ within the droplet at arbitrary moment of time $t$ at $t \geq t_0$. As follows from Eq. (4) and from Eqs. (12) and (11) with account of Eq. (9) (which allows us to express $D_2 [n_2 - n_{2\infty}(1 - c_{1s})]$ through $D_1 [n_1 - n_{1\infty}(c_{1s})]$),

$$x_1 + x_2 = \frac{8\pi}{3} Rt \frac{D_1}{c_{1s}} [n_1 - n_{1\infty}(c_{1s})] \frac{c_{1s} v_1 + (1 - c_{1s}) v_2}{c_1 v_1 + (1 - c_1) v_2}. \tag{15}$$

Using Eqs. (8) and taking into account Eqs. (9), (6), (7), and equality $c_2 - c_{2s} = -(c_1 - c_{1s})$, one can find

$$(1 - c_1) \frac{dx_1}{dt} - c_1 \frac{dx_2}{dt} = -4\pi (c_1 - c_{1s}) R \frac{D_1}{c_{1s}} [n_1 - n_{1\infty}(c_{1s})] - \\ - 4\pi (c_1 - c_{1s}) R \left[ (1 - c_1) D_1 (\tilde{n}_{1\infty} + f_1(c_1)) + c_1 D_2 (\tilde{n}_{2\infty} + f_2(c_1)) \right] \tag{16}$$

where

$$f_1(c_1) \equiv \frac{n_{1\infty}(c_1) - n_{1\infty}(c_{1s})}{c_1 - c_{1s}} - \tilde{n}_{1\infty} = \sum_{k=2} \alpha_k^{(1)} \frac{c_1^k - c_{1s}^k}{c_1 - c_{1s}}, \tag{17}$$

$$f_2(c_1) \equiv \frac{n_{2\infty}(c_2) - n_{2\infty}(c_{2s})}{c_2 - c_{2s}} - \tilde{n}_{2\infty} = \sum_{k=2} \alpha_k^{(2)} \frac{c_2^k - c_{2s}^k}{c_2 - c_{2s}} \tag{18}$$

are the non-ideal contributions to ratios $\dfrac{n_{1\infty}(c_1) - n_{1\infty}(c_{1s})}{c_1 - c_{1s}}$ and $\dfrac{n_{2\infty}(c_2) - n_{2\infty}(c_{2s})}{c_2 - c_{2s}}$.

Substituting Eqs. (15) and (16) in Eq. (2) and canceling common factors from both sides lead to the following relaxation differential equation for concentration of a solution within a growing droplet in the diffusion-controlled regime

$$\frac{dc_1}{dt} = -\frac{3(c_1 - c_{1s})}{2t} \times$$
$$\times \left\{ 1 + \frac{(1 - c_1) c_{1s} D_1 (\tilde{n}_{1\infty} + f_1(c_1)) + c_1 c_{1s} D_2 (\tilde{n}_{2\infty} + f_2(c_1))}{D_1 [n_1 - n_{1\infty}(c_{1s})]} \right\} \frac{c_1 v_1 + (1 - c_1) v_2}{c_{1s} v_1 + (1 - c_{1s}) v_2}. \tag{19}$$

The fact that equation (19) contains the factor $c_1 - c_{1s}$, is the principal one for the iteration method of solution Eq. (19) with respect to the deviation $c_1 - c_{1s}$. When $c_1$ is close to $c_{1s}$, we can substitute with a good accuracy $c_1$ by $c_{1s}$ in the factors after $c_1 - c_{1s}$ on the right hand side of Eq. (19). Then, solving equation (19) with initial condition

$$c_1 |_{t=t_0} = c_{10}, \tag{20}$$

we obtain

$$c_1 - c_{1s} = (c_{10} - c_{1s}) \left( \frac{t_0}{t} \right)^{3(1+\eta)/2} \quad (t \geq t_0) \tag{21}$$

where

$$\eta \equiv \frac{c_{1s} \left[ (1 - c_{1s}) D_1 (\tilde{n}_{1\infty} + f_1(c_{1s})) + c_{1s} D_2 (\tilde{n}_{2\infty} + f_2(c_{1s})) \right]}{D_1 [n_1 - n_{1\infty}(c_{1s})]}. \tag{22}$$

According to Eq. (22), the parameter $\eta$ depends not only on the thermodynamic ($\tilde{n}_{1\infty}$, $\tilde{n}_{2\infty}$) and kinetic ($D_1, D_2$) characteristics of condensing vapors, but also on the steady concentration $c_{1s}$ itself and the characteristics of the solution non-ideality ($f_1(c_{1s})$, $f_1(c_{1s})$) at this concentration. With the help of Eq. (9), one can rewrite Eq. (22) in the form that shows the symmetry of the parameter η with respect to the first and the second condensing components. As follows from Eqs. (22), (17), (18), and from the thermodynamic stability conditions of solutions (Landau, Lifshitz, 1976) and the conditions (5), $\eta > 0$. The smallness of the parameter η with respect to unity is not required.

Formula (21) expresses the power law of setting the steady concentration in the binary solution within a growing droplet. As is clear from the consideration above, not only the power law (12) of droplet radius growth comes in force after time $t_0$ since nucleation of markedly supercritical droplet occurs, but also the power law (21) of setting the steady concentration in the droplet takes place.

Taking into account Eq. (14), we can rewrite Eq. (21) as

$$c_1 - c_{1s} = (c_{10} - c_{1s})\left(\frac{R_0}{R}\right)^{3(1+\eta)} \quad (t \geq t_0). \tag{23}$$

Because $\eta > 0$, formula (23) shows that the steady concentration is established at comparatively small relative increase of droplet radius (and at very small relative increase of droplet radius, if inequality $\eta \geq 1$ holds).

Taking $R_0/\lambda \sim 10 \div 20$, $\lambda \sim 3 \cdot 10^{-5}$ cm, $D_1 \approx D_2 \sim 2 \cdot 10^{-1}$ см$^2 \cdot$с$^{-1}$, $v_1 \approx v_2 \sim 3 \cdot 10^{-23}$ см$^3$, $n_1 - n_{1\infty}(c_{1s}) \approx n_2 - n_{2\infty}(c_{2s}) \sim 10^{18}$ см$^{-3}$, we obtain with the help of Eqs. (13) and (11) an estimate for important time $t_0$ in the diffusion-controlled regime of droplet growth,

$$t_0 \sim \left(10^{-3} \div 10^{-2}\right) c. \tag{24}$$

**4. Free-molecular regime of droplet growth**

Let us consider now binary condensation in a free-molecular regime of droplet growth which realizes at $R/\lambda \ll 1$. As before, the relative amount of the passive gas in the vapor-gas environment is assumed to be a sufficiently large to provide an isothermal droplet growth and to fix λ as the free molecular path for vapor molecules in the passive gas. The path λ determines now the Knudsen layer width around the markedly supercritical droplet, i.e., the droplet with radius $R$ satisfying inequality $R \geq (3 \div 4) R_c$ where $R_c$ is the critical droplet radius, $R_c \ll \lambda$.

Instead of diffusion equations (8), we have now the free-molecular equations (Fuks, 1958) of droplet growth,

$$\frac{dx_1}{dt} = \pi \alpha_1 w_1 [n_1 - n_{1\infty}(c_1)] R^2, \qquad \frac{dx_2}{dt} = \pi \alpha_2 w_2 [n_2 - n_{2\infty}(c_2)] R^2 \qquad (25)$$

where $\alpha_1$ and $\alpha_2$ ($\alpha_1 \leq 1$, $\alpha_2 \leq 1$) are the condensation coefficients for molecules of the first and the second components of vapor mixture, $w_1$ and $w_2$ are the average thermal velocities of these molecules. As the implication of the Gibbs – Kelvin equation, the saturation concentrations $n_{1\infty}(c_1)$ and $n_{2\infty}(c_2)$ in Eqs. (25) do not depend on $R$ at $R \geq (3 \div 4) R_c$. Notice, that the Gibbs – Kelvin equation a fortiori provides also the absence of the dependence of $n_{1\infty}(c_1)$ and $n_{2\infty}(c_2)$ on $R$ in Eqs. (8) for the diffusion-controlled regime of droplet growth.

As follows from Eqs. (25) and (2), the steady concentration $c_{1s}$ satisfies now the equation

$$\frac{\alpha_1 w_1 [n_1 - n_{1\infty}(c_{1s})]}{\alpha_2 w_2 [n_2 - n_{2\infty}(1 - c_{1s})]} = \frac{c_{1s}}{1 - c_{1s}}. \qquad (26)$$

Differentiating both sides of Eq. (3) with respect to time $t$ at $c_1 = c_{1s}$ and fixed $v_1$ and $v_2$, using Eqs. (25) and integrating with respect to time, we obtain instead of Eq. (12) a linear-in-time law of the droplet radius growth in time,

$$R = \gamma t \qquad (\tau_0 \leq t \leq 3\tau_0) \qquad (27)$$

where

$$\gamma \equiv \frac{1}{4} \alpha_1 w_1 v_1 [n_1 - n_{1\infty}(c_{1s})] + \frac{1}{4} \alpha_2 w_2 v_2 [n_2 - n_{2\infty}(1 - c_{1s})], \qquad (28)$$

$$\tau_0 \equiv 30 R_c / \gamma. \qquad (29)$$

Usually, $R_c \sim 10^{-7} cm$ and $\lambda \sim 3 \cdot 10^{-5} cm$. In this case, the linear law (27) is valid with a high accuracy at $\tau_0 \leq t \leq 3\tau_0$. Indeed, according to Eqs. (27) and (29), we have $30 R_c \leq R \leq 90 R_c$. Thus the droplet was markedly supercritical long before establishing the linear law (27) (it explains the lower limitation $t \geq \tau_0$ for $t$), and the condition $R/\lambda \ll 1$ of a free-molecular regime of growth is still fulfilled after establishing the linear law (it explains the upper limitation $t \leq 3\tau_0$ for $t$). The following discussion in this Section will be referred to times $\tau_0 \leq t \leq 3\tau_0$. The positivity of the parameter $\gamma$ which follows from (28) and (5), corresponds to the droplet growth. Evidently one can rewrite Eq. (27) also as

$$R/R_0 = t/\tau_0 \quad (\tau_0 \leq t \leq 3\tau_0) \tag{30}$$

where

$$R_0 \equiv R\big|_{t=\tau_0} \tag{31}$$

(in the previous section $R_0$ was interpreted according to $R_0 \equiv R\big|_{t=t_0}$).

Let us formulate an iteration method of finding how the concentration $c_1$ within the growing droplet depends on time $t$ at the free-molecular regime of the droplet growth starting from time $\tau_0$. As follows from Eq. (4) and from Eqs. (27) and (28) with account of Eq. (26) (which allows us to express $\alpha_2 w_2 [n_2 - n_{2\infty}(1-c_{1s})]$ through $\alpha_1 w_1 [n_1 - n_{1\infty}(c_{1s})]$),

$$x_1 + x_2 = \frac{\pi}{3} R^2 t \frac{\alpha_1 w_1}{c_{1s}} [n_1 - n_{1\infty}(c_{1s})] \frac{c_{1s} v_1 + (1-c_{1s}) v_2}{c_1 v_1 + (1-c_1) v_2}. \tag{32}$$

Using equations (25) and taking into account Eqs. (26), (6), (7), and equality $c_2 - c_{2s} = -(c_1 - c_{1s})$, one can find

$$\begin{aligned}(1-c_1)\frac{dx_1}{dt} - c_1 \frac{dx_2}{dt} = &-\pi(c_1 - c_{1s}) R^2 \frac{\alpha_1 w_1}{c_{1s}} [n_1 - n_{1\infty}(c_{1s})] - \\ &-\pi(c_1 - c_{1s}) R^2 \left[(1-c_1)\alpha_1 w_1 (\tilde{n}_{1\infty} + f_1(c_1)) + c_1 \alpha_2 w_2 (\tilde{n}_{2\infty} + f_2(c_1))\right]\end{aligned} \tag{33}$$

where $f_1$ and $f_2$ are determined by Eqs. (17) and (18).

Substituting Eqs. (32) and (33) in Eq. (2) and canceling common factors from both sides lead to the relaxation differential equation for concentration of a solution within a growing droplet in the free-molecular regime

$$\frac{dc_1}{dt} = -\frac{3(c_1 - c_{1s})}{t} \times$$
$$\times \left\{ 1 + \frac{(1-c_1)c_{1s}\alpha_1 w_1 (\tilde{n}_{1\infty} + f_1(c_1)) + c_1 c_{1s}\alpha_2 w_2 (\tilde{n}_{2\infty} + f_2(c_1))}{\alpha_1 w_1 [n_1 - n_{1\infty}(c_{1s})]} \right\} \frac{c_1 v_1 + (1-c_1)v_2}{c_{1s} v_1 + (1-c_{1s})v_2}. \tag{34}$$

This equation is an analog of Eq. (19). Due to the factor $c_1 - c_{1s}$ on the right hand side of Eq. (34), we may apply the iteration method of its solution with respect to the deviation $c_1 - c_{1s}$. When $c_1$ is close to $c_{1s}$, we can substitute with a good accuracy $c_1$ by $c_{1s}$ in the factors after $c_1 - c_{1s}$ on the right hand side of Eq. (34). Then, solving equation (34) with initial condition

$$c_1|_{t=\tau_0} = c_{10}, \tag{35}$$

we obtain the power time law for setting the steady concentration in the free-molecular regime of droplet growth,

$$c_1 - c_{1s} = (c_{10} - c_{1s})\left(\frac{\tau_0}{t}\right)^{3(1+\chi)} \quad (\tau_0 \leq t \leq 3\tau_0). \tag{36}$$

The parameter $\chi$ is determined here by the right-hand side of Eq. (22) with substitution of $D_1$ and $D_2$ by $\alpha_1 w_1$ and $\alpha_2 w_2$, i.e.

$$\chi \equiv \frac{c_{1s}\left[(1-c_{1s})\alpha_1 w_1(\tilde{n}_{1\infty} + f_1(c_{1s})) + c_{1s}\alpha_2 w_2(\tilde{n}_{1\infty} + f_2(c_{1s}))\right]}{\alpha_1 w_1 [n_1 - n_{1\infty}(c_{1s})]}. \tag{37}$$

According to (37), the parameter $\chi$ depends not only on thermodynamic ($\tilde{n}_{1\infty}$, $\tilde{n}_{2\infty}$) and kinetic ($\alpha_1 w_1$, $\alpha_2 w_2$) characteristics of condensing vapors, but also on the steady concentration $c_{1s}$ itself and the characteristics of the solution non-ideality ($f_1(c_{1s})$, $f_1(c_{1s})$) at this concentration. As follows from Eqs. (37), (17), (18), and from the thermodynamic stability conditions of solutions (Landau, Lifshitz, 1976) and the conditions (5), $\chi > 0$. The smallness of the parameter $\chi$ with respect to unity is not required.

It is clear that the power law (27) of droplet radius growth and the power law (36) of setting steady concentration in the solution within the droplet are valid in the same range $\tau_0 \leq t \leq 3\tau_0$ of times $t$ (counted off the moment of nucleation of a markedly supercritical droplet in vapor-gas environment).

Taking into account Eq. (30), we can rewrite Eq. (36) as

$$c_1 - c_{1s} = (c_{10} - c_{1s})\left(\frac{R_0}{R}\right)^{3(1+\chi)} \quad (\tau_0 \leq t \leq 3\tau_0). \tag{38}$$

Since $\chi > 0$, formula (38) shows that the steady concentration of the solution within the droplet establishes at comparatively small relative increase of droplet radius (and even at very small relative increase of droplet radius, if inequality $\chi \geq 1$ takes place).

Let us estimate important time $\tau_0$ with the help of Eqs. (28) and (29). Taking $R_c \sim 10^{-7}$ см, $\alpha_1 \approx \alpha_2 \sim 1$, $w_1 \approx w_2 \sim 5 \cdot 10^4$ см×с$^{-1}$, $v_1 \approx v_2 \sim 3 \cdot 10^{-23}$ см$^3$, $n_1 - n_{1\infty}(c_{1s}) \approx n_2 - n_{2\infty}(c_{2s}) \sim 10^{18}$ см$^{-3}$, we obtain

$$\tau_0 \sim 10^{-5} c. \tag{39}$$

Comparing the estimate (39) with the estimate (24), we conclude that $\tau_0 \ll t_0$.

## 5. Total time dependence of concentration of a solution within a growing droplet with ideal solution

Whether the setting of a steady concentration in solution within a growing droplet will be able at any initial concentration $c_{10}$ in initial conditions (20) and (35) and at any values of thermodynamic and kinetic characteristics of condensing components, and will occur for a sufficiently short time allowing us to consider the droplet growth to be steady, is an interesting question. An answer to this question can not be obtain within the frameworks of the theory considered in Sections 3 and 4 where the relaxation equations have been solved under a priori assumption that $c_1$ is close to $c_{1s}$. However, solving the general relaxation equations (19) and (34) without the assumption of a smallness of the deviation of $c_1$ from $c_{1s}$, we may find the limitations on the initial concentration $c_{10}$ in initial conditions (20) and (35) and the limitations on the values of thermodynamic and kinetic characteristics of condensing components, at which the conclusions made in Sections 3 and 4 are confirmed or violated. In this way we will find an answer to the question posed above.

General relaxation equations (19) and (34) may be strictly rewritten at ideality of solution (when Eq. (6) holds) in the form

$$\frac{d\delta c_1}{dt} = -\frac{\gamma}{t}\delta c_1 (1+a\delta c_1)(1+b\delta c_1) \qquad (40)$$

where

$$\delta c_1 \equiv c_1 - c_{1s}, \qquad \delta c_{10} \equiv c_{10} - c_{1s}, \qquad (41)$$

$$\gamma \equiv \begin{cases} \dfrac{3}{2}(1+\eta) & \text{(diffusion-controlled regime)}, \\ \\ 3(1+\chi) & \text{(free-molecular regime)}, \end{cases} \qquad \gamma > 0, \qquad (42)$$

$$a \equiv \begin{cases} -\dfrac{c_{1s}}{1+\eta}\dfrac{D_1\tilde{n}_{1\infty} - D_2\tilde{n}_{2\infty}}{D_1[n_1 - n_{1\infty}(c_{1s})]} & \text{(diffusion-controlled regime)}, \\ \\ -\dfrac{c_{1s}}{1+\chi}\dfrac{\alpha_1 w_1 \tilde{n}_{1\infty} - \alpha_2 w_2 \tilde{n}_{2\infty}}{\alpha_1 w_1[n_1 - n_{1\infty}(c_{1s})]} & \text{(free-molecular regime)}, \end{cases} \qquad (43)$$

$$b \equiv \frac{v_1 - v_2}{c_{1s}v_1 + (1-c_{1s})v_2}. \qquad (44)$$

The values of $\delta c_1$ and coefficients $a$ and $b$ can be of any sign.

An analytical solution of equation (40) with initial conditions (20) and (35) can be found by separating variables and decomposition in partial fractions. This solution has a form

$$\gamma \ln\left(\frac{t}{t_i}\right) = -\ln\left(\frac{\delta c_1(t)}{\delta c_{10}}\right) + \frac{a}{a-b}\ln\left(\frac{1+a\delta c_1(t)}{1+a\delta c_{10}}\right) - \frac{b}{a-b}\ln\left(\frac{1+b\delta c_1(t)}{1+b\delta c_{10}}\right) \quad (45)$$

where we denoted for brevity $t_0$ and $\tau_0$ as single time $t_i$. The solution exists (within the real-valued domain) only when the following conditions fulfill

$$\frac{\delta c_1}{\delta c_{10}} > 0, \quad \frac{1+a\delta c_1}{1+a\delta c_{10}} > 0, \quad \frac{1+b\delta c_1}{1+b\delta c_{10}} > 0. \quad (46)$$

The solution (45) can be rewritten equivalently with account of Eq. (42) as

$$\frac{c_1-c_{1s}}{c_{10}-c_{1s}}\left[\frac{1+a(c_{10}-c_{1s})}{1+a(c_1-c_{1s})}\right]^{\frac{a}{a-b}}\left[\frac{1+b(c_1-c_{1s})}{1+b(c_{10}-c_{1s})}\right]^{\frac{b}{a-b}} =$$

$$= \begin{cases} \left(\dfrac{t_0}{t}\right)^{3(1+\eta)/2} & \text{(diffusion-controlled regime)}, \\[2mm] \left(\dfrac{\tau_0}{t}\right)^{3(1+\chi)} & \text{(free-molecular regime)}. \end{cases} \quad (47)$$

Below we will suppose that $\delta c_{10} > 0$, and, in view of Eq. (46), also $\delta c_1 > 0$. In the case when $\delta c_1 < 0$, we have from Eq. (40) an equation for $|\delta c_1| \equiv -\delta c_1$, which differs from Eq. (40) only in replacement of $\delta c_1$, $1+a\delta c_1$, and $1+b\delta c_1$ by $|\delta c_1|$, $1-a|\delta c_1|$, and $1-b|\delta c_1|$, respectively, i.e. it is equivalent to the replacement of the values of coefficients $a$ and $b$ by values $-a$ and $-b$. As is clearly seen, solution (47) transforms into solutions (21) and (36) when $c_{10}$ and, in view of Eq. (40), $c_1$ are close to $c_{1s}$ or when $|a| \ll 1$ and $|b| \ll 1$. Moreover, we may ascertain from Eqs. (46) и (47) with the help of Eq. (40) that positive ratio $(c_1-c_{1s})/(c_{10}-c_{1s})$ decreases monotonously to zero with increasing $t$ at $|a|<1$, $|b|<1$, and at any value of initial concentration $c_{1s} < c_{10} < 1$. The same behavior will be at any positive values of coefficients $a$ and $b$. Thus we have in both last cases a confirmative answer to the question posed at the beginning of this Section.

However, it follows from Eq. (40) that there is no monotonous decrease to zero for solution (47) with increasing time $t$ at $a>0$, $b<0$, $|b|>1$, if initial concentration satisfies inequality $c_{1s}+1/|b|<c_{10}<1$. Note, at this $d\delta c_1/dt|_{t=t_i}>0$. Such initial concentration may be realized in practice only when $0<c_{1s}\ll 1$ or $|b|\gg 1$. It also follows from Eq. (40) that a monotonous decrease to zero for solution (47) with increasing time $t$ is impossible at $a<0$, $b<0$, $|a|<1$, and $|b|>1$, as well as at $a<0$, $b<0$, $|a|>1$, $|b|>1$, and $|b|<|a|$, if initial concentration satisfies the previous inequality $c_{1s}+1/|b|<c_{10}<1$. Note here, that the first situation corresponds to $d\delta c_1/dt|_{t=t_i}>0$, while the second situation corresponds to $d\delta c_1/dt|_{t=t_i}<0$ with $\delta c_1 \to 1/|b|$ as $t\to\infty$. All what was said above in this paragraph is true if we exchange the values of parameters $a$ and $b$.

Thus we see that some initial values $c_{10}$ should be excluded at negative values of coefficients $a$ and $b$. In this way, we give a full answer to the question posed at the beginning of this Section.

## 6. Discussion and conclusions

We have considered reaching the power-law-in-time growth of the droplet radius and setting the steady concentration in the binary solution within the growing droplet at two characteristic regimes of isothermal binary condensation in a vapor-gas environment. In fact, the regimes of droplet growth can gradually change from the free-molecular to the diffusion-controlled one with the increase of droplet size. Evidently, the transition from Eq. (9) for the steady concentration of solution in the droplet at the diffusion-controlled regime to Eq. (26) for the steady concentration of solution in the droplet at the free-molecular regime requires formally a replacement of the ratio $D_1/D_2$ by the ratio $\alpha_1 w_1/\alpha_2 w_2$. At the typical proximity of the order of magnitude of the ratios $D_1/D_2$ and $\alpha_1 w_1 / \alpha_2 w_2$, Eqs (9) and (26) are almost equivalent. Hence, the steady concentrations in two considered regimes should be close also.

Similarly, the transition from Eq. (22) for the parameter $\eta$ in the power law (21) of setting the steady concentration in the case of the diffusion-controlled regime to Eq (37) for the parameter $\chi$ in the power law (36) of setting the steady concentration in the case of the free-molecular regime requires formally a replacement of the ratio $D_1/D_2$ by the ratio $\alpha_1 w_1 / \alpha_2 w_2$. At the typical proximity of the order of magnitude of the relations $D_1/D_2$ and $\alpha_1 w_1 / \alpha_2 w_2$, the laws (21) and (36) of steady concentration setting in two considered regimes would be then approximately similar (at the time limitations noted in the laws).

As the droplet grows in the diffusion-controlled regime, the condition $R/\lambda \gg 1$ of this regime holds better and better. Vise versa, as the droplet grows in the free-molecular regime, the condition $R/\lambda \ll 1$ of that regime holds worse and worse. This fact results in large difference in the time intervals available for the theory in both considered cases.

The number densities $\tilde{n}_{1\infty}$ and $\tilde{n}_{2\infty}$ of the molecules of saturated vapors of pure liquids can strongly differ in their values (for instance, it is true for the vapors of water and sulphuric acid). The analytical formula for binary solution steady concentration obtained by Kulmala et al. (1993) can give at that very different particular values of the steady concentration $c_{1s}$. Because the parameters $\eta$ and $\chi$ in the power laws (21) and (36) depend, according to Eqs. (22) and (37), on both densities $\tilde{n}_{1\infty}$ and $\tilde{n}_{2\infty}$, and on the steady concentration $c_{1s}$, then the values of the positive parameters $\eta$ and $\chi$ can lie in a very wide range. In particular, if the relations $\eta \geq 1$ and $\chi \geq 1$ take place, then, according to Eqs. (23) and (38), a steady concentration of droplet solution would establish at very small relative increase of droplet radius. It means that the droplet would grow at the diffusion-controlled and the free-molecular regimes at steady concentration of the droplet solution for a major period of time. Then simple formulas (12) and (27) are valid for the time dependence of a droplet radius.

The results obtained in Section 5 confirm the results of Sections 3 and 4 concerned with the monotonous setting of the steady concentration in a droplet and found by solving relaxation equations (19) and (34) with a priori assumption of proximity of the droplet concentration to the steady concentration.

**Acknowledgments**

This work was supported financially by the program of Federal Agency of Education "The development of scientific potential of high school (2006 – 2008)" (project RNP.2.2.2.1712). The authors thank Dr. D.V. Tatyanenko for the discussion of the work and useful remarks.